\documentclass[twocolumn,showpacs,preprintnumbers,amsmath,amssymb]{revtex4}

\usepackage{graphicx}

\begin{document}

\title{Elliptical hole pockets in the Fermi surfaces of unhydrated and hydrated sodium cobalt oxides}

\author{J. Laverock and S.~B. Dugdale}
\affiliation{H.~H.~Wills Physics Laboratory, University of Bristol, Tyndall Avenue, Bristol BS8 1TL, United Kingdom}

\author{J.~A. Duffy, J. Wooldridge, G. Balakrishnan and M.~R. Lees}
\affiliation{Department of Physics, University of Warwick, Coventry CV4 7AL, United Kingdom}

\author{G.-q. Zheng}
\affiliation{Department of Physics, Okayama University, Okayama 700-8530, Japan}

\author{D. Chen and C.~T. Lin}
\affiliation{ISEM, University of Wollongong, Wollongong, NSW 2522, Australia}

\author{A. Andrejczuk, M. Itou and Y. Sakurai}
\affiliation{Japan Synchrotron Radiation Research Institute, SPring-8, 1-1-1 Kouto, Mikazuki, Sayo, Hyogo 679-5198, Japan}

\begin{abstract}
The surprise discovery of superconductivity below 5K in sodium cobalt oxides when hydrated with water has
caught the attention of experimentalists and theorists alike. Most explanations for its occurence
have focused heavily on the properties of
some small elliptically shaped pockets predicted to be the electronically dominant Fermi surface sheet,
but direct attempts to look for them have instead
cast serious doubts over their existence. Here we present evidence that these pockets do indeed exist,
based on bulk measurements of the electron momentum distribution in unhydrated and hydrated sodium cobalt
oxides using the technique of x-ray Compton scattering.
\end{abstract}

\pacs{71.18.+y,71.27.+a,74.70.-b}

\maketitle

Hydrated sodium cobalt oxides (Na$_{x}$CoO$_{2} \cdot $1.3H$_{2}$O), for a
 certain range of Na concentrations, exhibit
superconductivity at a temperature
of 5K \cite{takada2003}.
Although many analogies have been drawn with the high-$T_{c}$ cuprates
(for instance, as being possibly the only other example of a Mott
insulator becoming superconducting under doping \cite{takada2003,wang2004b}), the
sodium cobaltate system exhibits its own unique set of anomalous behaviour
such as unusually high thermopower \cite{terasaki1997} and $T$-linear
resistivity \cite{wang2003}), distinctly indicative of strongly
correlated electron behaviour.  In a conventional
superconductor, electrons at the Fermi surface form Cooper pairs under an attractive interaction mediated by lattice vibrations. The 
manner in which electrons form these pairs can be strongly influenced by the shape of the
Fermi surface.  Questions regarding
the origin of the pairing interaction and the nature of the superconductivity in the cobaltates has
stimulated significant theoretical speculation, most of which has focused heavily on the properties of
some small elliptically shaped pockets predicted to be the electronically dominant Fermi surface sheet 
\cite{johannes2004b,kuroki2004,mochizuki2005a,mazin2005},
but the outcome of direct attempts to look for them has instead
cast serious doubt over their existence \cite{hasan2004,yang2004,yang2005,qian2006a,qian2006b}.
Here we present evidence that these pockets do indeed exist, 
based on bulk measurements of the electron momentum distribution in unhydrated and hydrated sodium cobalt
oxides using the technique of x-ray Compton scattering.

The structure of Na$_{x}$CoO$_{2}$ comprises
hexagonal planes of electronically active edge-sharing CoO$_{6}$ octahedra \cite{viciu2006}.
These planes are separated by insulating layers of Na, and, in the hydrated samples,
 water, that serve as spacers (resulting in electronic two-dimensionality) and charge
reservoirs.  The intercalation of water, at a concentration of $y \sim 1.3$, has a dramatic effect
on properties of the compound.  It is accompanied by a near doubling of the $c$-axis lattice parameter and,
for $x \sim 0.3$, the onset of superconductivity.
Calculations of electronic structure based on the local-density-approximation (LDA)
predict, for concentrations $x \leq \sim$0.6 in Na$_{x}$CoO$_{2}$,  a Fermi surface 
composed of two hole sheets originating  
from Co $d$-orbitals \cite{singh2000,zhang2004,zhou2005}.
The larger sheet, of $a_{1g}$ character, is centred at the $\Gamma$-point
(zone centre) whereas the other sheet, of  $e^{\prime}_{g}$ character, comprises 
six smaller elliptical cylinders located between $\Gamma$ and
$K$ (for $x \geq 0.7$, this band becomes completely filled and does not
contribute to the Fermi surface).
These $e^{\prime}_{g}$ pockets may be of considerable importance, since many of the models
for superconductivity in Na$_{x}$CoO$_{2} \cdot y$H$_{2}$O are predicated upon their existence. 
In particular, an analysis of permitted order-parameter symmetries has indicated that
an unconventional $f$-wave ($x(x^{2}-3y^{2}) {\hat{\bf z}}$) would be the most probable
if the pockets exist \cite{mazin2005}. 
With 70\% of the density-of-states at the
Fermi level, these pockets are central to several models of
spin-fluctuation-mediated superconductivity, developing either via imperfect
nesting between the Fermi surface pockets \cite{johannes2004b}, or as a consequence
of these disconnected pieces of Fermi surface \cite{kuroki2004,mochizuki2005a}.

A series of angle-resolved photoemission
spectroscopy (ARPES) results, the consensus of which report observation of the $a_{1g}$ sheet at $\Gamma$ but
 crucially not the $e^{\prime}_{g}$ pockets, has fuelled the controversy surrounding the elliptical pockets. Indeed, they
observe the $e^{\prime}_{g}$ band as lying completely below the Fermi level
\cite{yang2004,yang2005,qian2006a,qian2006b}. However, it is well-known that surface effects and matrix elements
can have a strong influence on photoemission results, although there are recent reports of measurements being made with
an electron escape depth of 200 \AA \cite{shimojima2006}. 
Here we directly tackle this controversy by presenting a Compton scattering study both of the Fermi surfaces of a representative
set of compositions of the unhydrated parent compound Na$_{x}$CoO$_{2}$ ($x$=0.38, 0.51 and 0.74)
and a hydrated (actually deuterated) sample at a superconducting composition, Na$_{0.35}$CoO$_{2} \cdot 1.3$D$_{2}$O.
For these measurements, single crystals of $x \sim 0.75$ were grown in
Warwick using the
floating-zone technique.  Samples of lower sodium concentration were then
obtained by a chemical deintercalation method, immersing the crystals
in solutions of Br$_2$ and acetonitrile.  The lattice parameters were
obtained by x-ray diffraction; the relationship between Na doping and the
crystal structure is well characterised by powder neutron diffraction
measurements and ICP-AES techniques \cite{huang2004c} and so the
sodium concentrations for the three crystals used in this study were
determined as 0.74(1), 0.51(1) and 0.38(1). The macroscopic properties
(magnetic susceptibility, heat capacity and transport measurements 
\cite{wooldridge2005})
are identical to those previously reported for similar compositions
\cite{chou2004b}. The superconducting sample was produced by
the chemical intercalation of deuterium oxide (D$_{2}$O was used rather than
H$_{2}$O to allow future neutron experiments on the same sample) by submersion
in liquid D$_{2}$O for three months at 5$^{\circ}$C \cite{chou2004a}. Subsequent to
the Compton experiment, a measurement of its magnetisation showed that it was superconducting at
a temperature of 3.5K.

\begin{figure}[t!]
\begin{center}
\includegraphics[width=60ex]{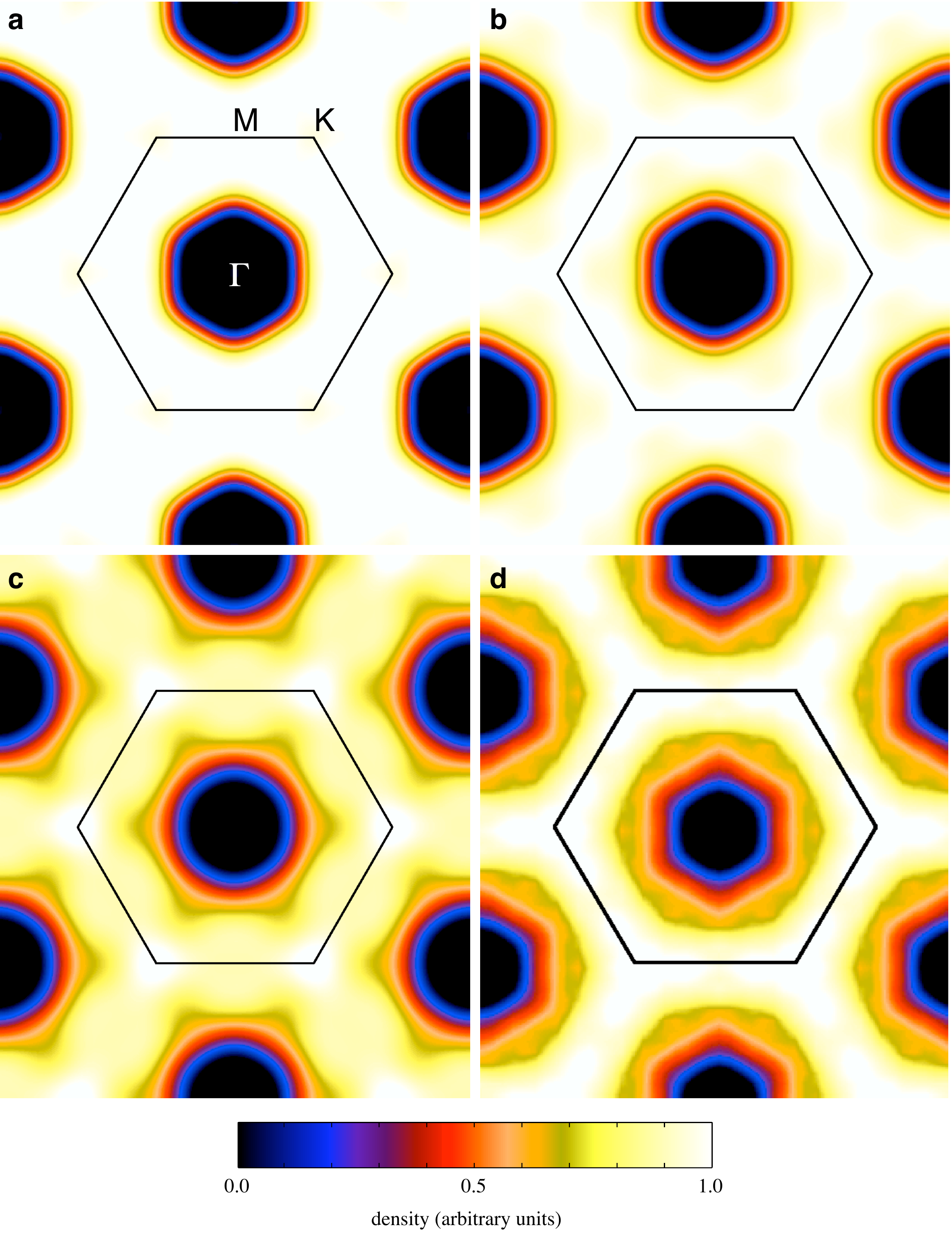}
\end{center}
\caption{\label{expfs} (Color online) The experimental Fermi surface of Na$_{x}$CoO$_{2}$ for
(a) $x=0.74$, (b) 0.51, and (c) 0.38,  and for Na$_{0.35}$CoO$_{2} \cdot 1.3$D$_{2}$O (d)
obtained from the reconstruction of five Compton
profiles for each composition. The boundary of the first Brillouin zone is indicated.}
\end{figure}

A Compton profile represents a double integral (one-dimensional projection)
of the full three-dimensional electron momentum density.
For each composition, five Compton profiles equally spaced between
$\Gamma$-$M$ and $\Gamma$-$K$ were measured on the
high-resolution Compton spectrometer of beamline BL08W at the SPring-8
synchrotron. The unhydrated measurements were made at room temperature,
while those on the hydrated sample were at 11K.  The hydrated sample was
transferred from a D$_{2}$O bath directly onto a cryostat precooled to $\sim 250$K under a helium atmosphere
in order to preserve the hydration.  On removal at the end of the measurement, the sample
was observed to be still in its hydrated state. The spectrometer consists of a
Cauchois-type crystal analyser and a position-sensitive detector,
with a resolution FWHM at the Compton peak of 0.115~a.u (1~a.u.\ of
momentum~$=$~1.99~$\times$10$^{-24}$~kg~m~s$^{-1}$)
\cite{hiraoka2001,sakurai2004}.
For each Compton profile, $\sim 600~000$ counts in the peak data channel
were accumulated, and each Compton profile was corrected for possible
multiple-scattering contributions.
A two-dimensional momentum density, representing a projection down the
$c$-axis of the full three-dimensional density, was reconstructed from
each set of five profiles using tomographic techniques \cite{kontrymsznajd1990}
and then folded back into the first BZ using the Lock-Crisp-West procedure
\cite{major2004,dugdale2006} to obtain the occupation density from which the occupied
parts of the BZ could be inferred.
The occupation density is shown for each composition in Fig.~\ref{expfs}, where black represents the
lowest occupancy and white the highest.

Considering first the unhydrated parent compound, the contours associated with the
hexagonal $a_{1g}$ hole sheet can be clearly identified for $x=0.74$,  but the hexagonal
shape becomes progressively less clear and is significantly distorted in the $x=0.38$ data. We shall argue that
this distortion is strong evidence for the presence of the $e^{\prime}_{g}$ pockets.

\begin{figure}[t!]
\begin{center}
\includegraphics[width=36ex,angle=90]{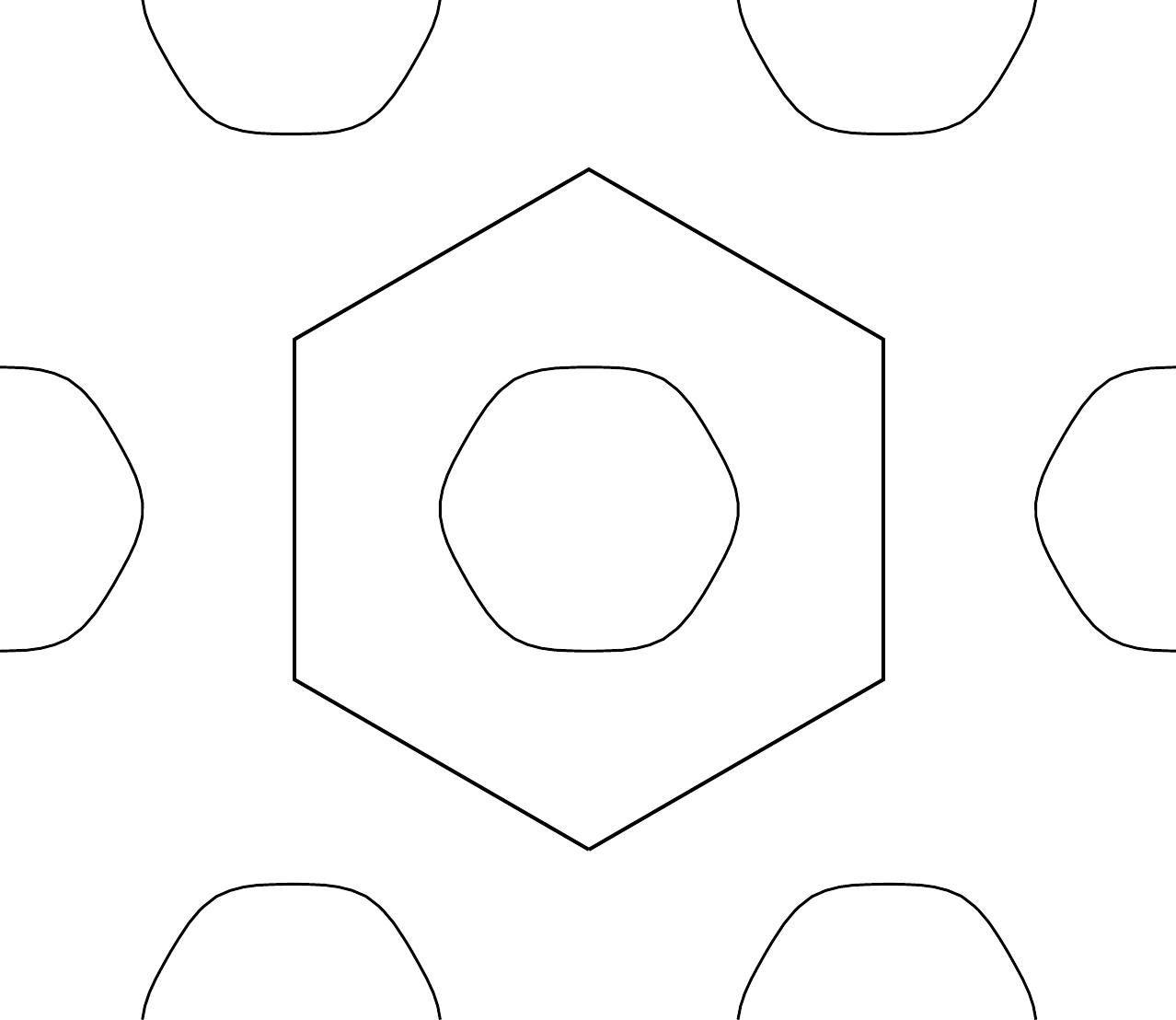}
\end{center}
\caption{\label{fderiv} Fermi surface obtained by plotting the contour at the maximum of the first derivative of the occupation density
 for Na$_{0.74}$CoO$_{2}$.}
\end{figure}

To assess the size of the hexagonal Fermi surface of the Na$_{0.74}$CoO$_{2}$ compound,
 a method using the extrema in the first derivative of the occupation density was employed
(see for example \cite{major2004}). At
this composition elliptical pockets are not expected, and so the determination using this method should unambiguously reveal
the hexagonal Fermi surface. Fig.~\ref{fderiv} is the result, and shows a Fermi surface in excellent agreement 
with LDA calculations.

\begin{figure}[t!]
\begin{center}
\includegraphics[width=36ex,angle=90]{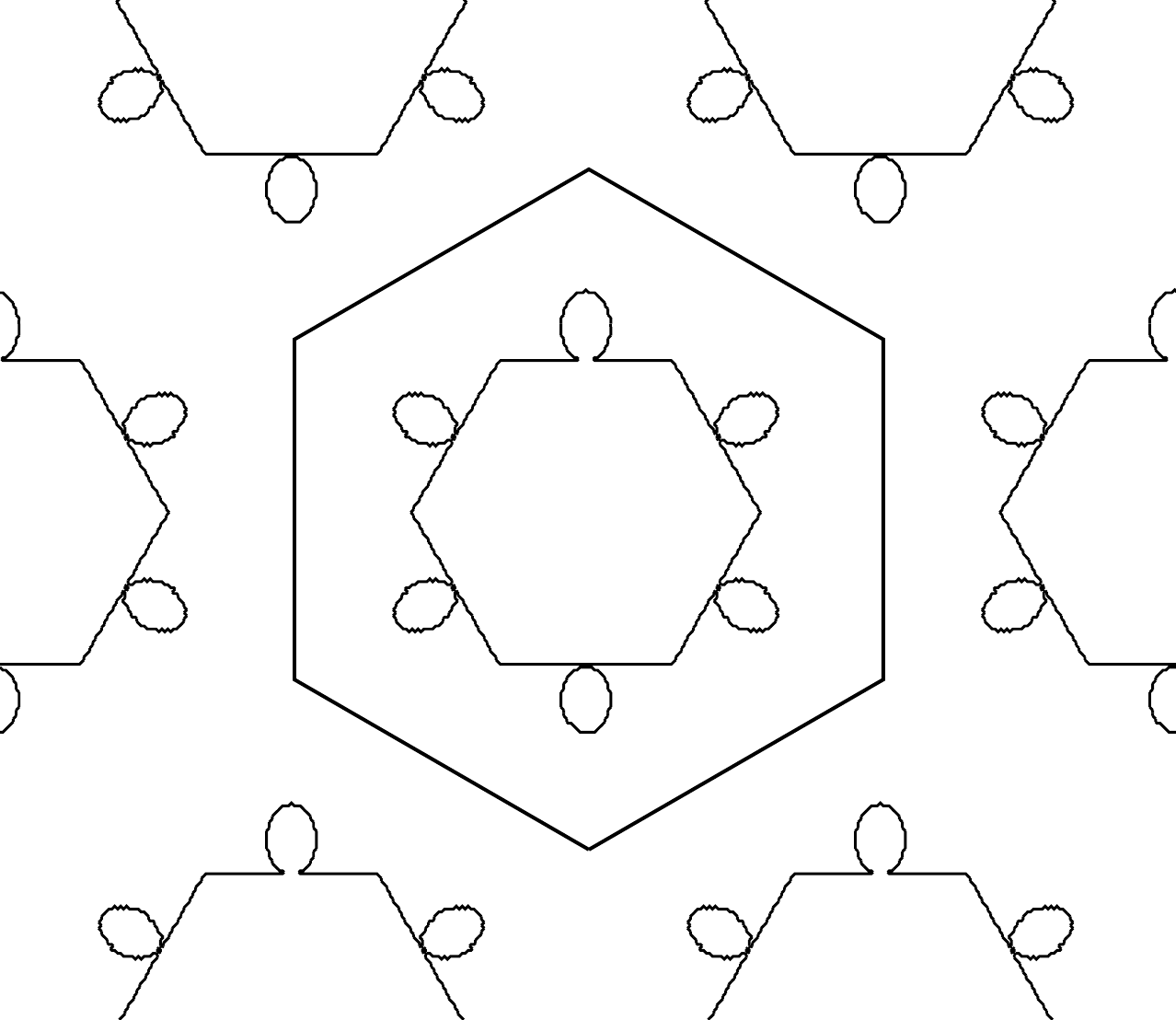}
\includegraphics[width=36ex,angle=90]{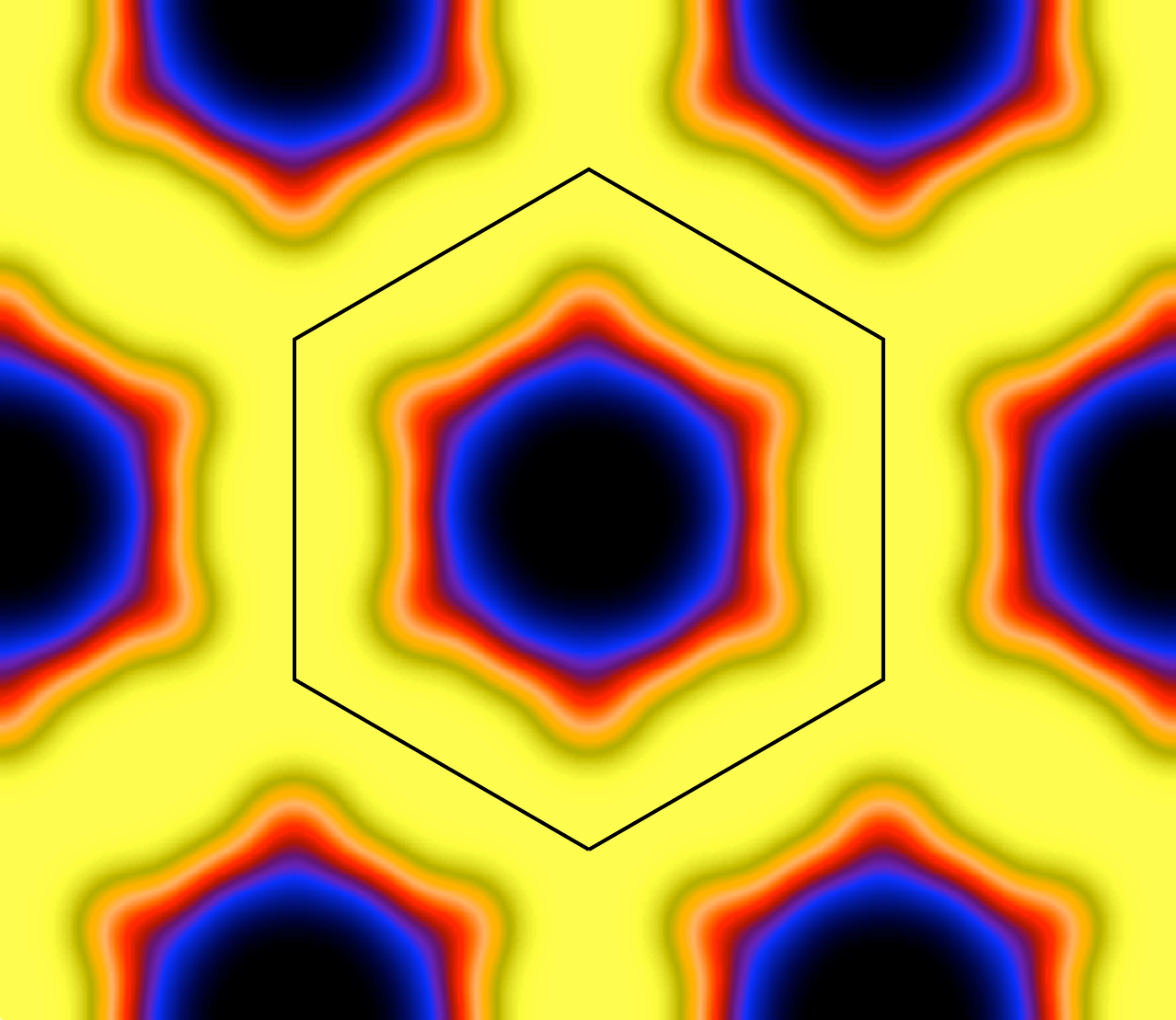}
\end{center}
\caption{\label{simul} (Color online) A simulation of a Fermi surface comprising a central hexagonal sheet (representing the $a_{1g}$ hole
sheet) and six $e^{\prime}_{g}$ elliptical hole pockets (left) together with the resulting occupation density convoluted with
the experimental resolution (right). }
\end{figure}

We can explain the distortion of the hexagonal shape for smaller Na concentrations 
as being due to the presence of $e^{\prime}_{g}$ elliptical hole pockets close to the central
$a_{1g}$ Fermi surface. A simple geometric simulation of such a Fermi surface is shown in Fig.~\ref{simul}
together with the resulting occupation density (convoluted with the experimental resolution), illustrating how
the presence of small pockets distort the hexagonal appearance of the $a_{1g}$ Fermi surface. The experimental
occupation density for the hydrated Na$_{0.35}$CoO$_{2} \cdot 1.3$D$_{2}$O (Fig.~\ref{expfs}) is also strongly
suggestive of the presence of small hole pockets, which can be discerned close to the central hexagon; it is also worth
remarking that the $a_{1g}$ sheet retains a strong hexagonal shape, whereas electronic structure calculations
in the hydrated structure (although without the actual presence of water) predict something more circular \cite{johannes2004a}. 
Our results suggest that the effects of hydration on the Fermi surface are in fact rather modest, and perhaps not as
drastic as suggested by Xiao {\it et al.} \cite{xiao2006}.
An estimate of the areas of the hexagonal $a_{1g}$ sheet
and (where appropriate) the six elliptical $e^{\prime}_{g}$ pockets based on a comparison of simulations
to the experimental data (with the total area constrained by the appropriate Na concentration) is presented in Table~\ref{fsparam}.
That the pockets consistently appear rather close to the $a_{1g}$ sheet is also noteworthy.

\begin{table}[t!]
\vspace*{0.2in}
\begin{center}
\begin{tabular}{|c|c|c|}
\hline
Cobaltate	& Area ($a_{1g}$) 	& Area (e$^{\prime}_{g}$)	 \\
\hline
Na$_{0.38}$CoO$_{2}$ & $0.262$ & $0.048$  \\
Na$_{0.51}$CoO$_{2}$ & $0.218$ & $0.027$  \\
Na$_{0.74}$CoO$_{2}$ & $0.132$ & not seen  \\
Na$_{0.35}$CoO$_{2} \cdot$ 1.3 D$_{2}$O & $0.310$ & $0.016$  \\
\hline
\end{tabular}
\caption{
Estimate of areas of the $a_{1g}$ and e$^{\prime}_{g}$ Fermi surfaces (in the case of the latter, this is the total area of
all six pockets) based on applying the simulation of the Fermi surface as
illustrated in Fig.~\ref{simul} for the compositions studied, as a proportion of the hexagonal first Brillouin zone.}
\label{fsparam}
\end{center}
\vspace*{-0.2in}
\end{table}

Recent measurement of Shubnikov-de Haas oscillations in Na$_{0.3}$CoO$_{2}$ indicate the
presence of some unidentified Fermi surface pockets occupying approximately 0.6\% and 1.4\% of
the BZ \cite{balicas2006}, which is consistent
with our estimate of the size of each pocket we observe occupying about 0.8\% (Table~\ref{fsparam}).
In addition, an examination of phonon softening in this system by Rueff {\it et al.} \cite{rueff2006} is interpreted by
those authors as strong evidence for the existence of nested pockets.

The question of why the ARPES experiments have consistently not observed these
e$^{\prime}_{g}$ pockets remains. 
Issues such as surface sensitivity, including possible surface 
relaxations of CoO$_{6}$ octahedral contractions that destroy the pockets \cite{mochizuki2005b,mochizuki2006},
as well as matrix-element effects or Na disorder \cite{singh2006} must be possibilities. However, at least in the superconducting
compound, it is very difficult to reconcile the observed behaviour of the specific heat, or even understand
the presence of superconductivity without the presence of the $e^{\prime}_{g}$ pockets \cite{oeschler2005}.
Moreover, in an attempt to take into account Coulomb correlations not included in
the LDA, calculations \cite{zhang2004,zhou2005} based on the LDA$+U$ approach,
have suggested that for a sufficiently large Coulomb energy ($U > 3$ eV) 
the $e^{\prime}_{g}$ band is pulled below
the Fermi level, eliminating these smaller Fermi surface pockets for all Na concentrations.
However, other studies have put an upper limit of about 2.3eV on $U$
\cite{lee2005}, and when dynamical Coulomb correlations are incorporated
the effect is to stabilize the $e^{\prime}_{g}$ pockets \cite{ishida2005}.
The theoretical debate rages on, with predictions in support of \cite{ishida2007} and contrary to 
\cite{bourgeois2007} the existence of these pockets.

In conclusion, we have presented the Fermi surface of several members ($x=0.38$, 0.51
and 0.74) of the unhydrated sodium cobalt oxide Na$_{x}$CoO$_{2}$ and
of a hydrated composition Na$_{0.35}$CoO$_{2} \cdot 1.3$D$_{2}$O.
Reasonable {\it qualitative} agreement is observed between our
experimentally determined Fermi surfaces and the LDA predictions, and
there is clear evidence for the smaller e$^{\prime}_{g}$ elliptical hole
pockets which develop at lower Na concentrations than Na$_{0.74}$CoO$_{2}$. 
For Na$_{0.38}$CoO$_{2}$, their presence is clearly indicated in experimental maps of
the occupancy within the Brillouin zone. Most importantly, however, the occupancy map for
Na$_{0.35}$CoO$_{2} \cdot 1.3$D$_{2}$O also shows the presence of small e$^{\prime}_{g}$ elliptical hole
pockets.  While alternative models that describe the superconductivity arising as a consequence
of frustration on the triangular lattice \cite{kumar2003,ogata2003},
nesting across the large, hexagonal sheet \cite{tanaka2003}, or spin
fluctuations enhanced by the dopant dynamics \cite{baskaran2003} are not (and cannot be) ruled out, the
observation of the pockets lends strong support to theories based on their special nesting properties. 

\section*{Acknowledgements}
We acknowledge the financial support of the Royal Society (S.B.D.) and
the UK EPSRC, and invaluable discussions with Igor Mazin, Michelle Johannes and Zahid Hasan. 
This experiment was performed with the approval of the Japan
Synchrotron Radiation Research Institute (JASRI, proposal nos. 2005A0092-ND3a-np and 2005B0182-ND3a-np).
This work was partially supported by a Grant-in-Aid for Scientific Research (No.18340111) from the Ministry of Education, 
Culture, Sports, Science and Technology, Japan.

\end{document}